\documentclass[pdflatex,sn-mathphys-ay,referee]{sn-jnl}% Math and Physical Sciences Author Year Reference Style
\usepackage{fourier}
\usepackage{graphicx}%
\usepackage{amsmath,amssymb,amsfonts}%
\usepackage{cuted}% full-width equations in two-column mode; replaces REVTeX widetext
\usepackage{rotating}% sideways supplementary condition-log table
\usepackage{bm}%
\usepackage{booktabs}%
\usepackage{array}%
\usepackage{siunitx}%
\usepackage{xcolor}%

\hypersetup{colorlinks=true,linkcolor=blue,citecolor=blue,urlcolor=blue}
\newcommand{\maybeincludegraphics}[2][]{%
  \IfFileExists{#2}{\includegraphics[#1]{#2}}{%
    \fbox{\begin{minipage}[c][0.25\textheight][c]{0.88\linewidth}\centering\small Missing figure file:\\[0.5em]\texttt{\detokenize{#2}}\end{minipage}}%
  }%
}

\begin{document}
\title[Acoustic control of liquid-jet breakup]{Acoustic control of Rayleigh-regime liquid-jet breakup using a through-hole MHz transducer}

\author[1]{\fnm{K.} \sur{Nguyen}}

\author[2]{\fnm{K.} \sur{Ahmed}}

\author*[1,3]{\fnm{J.} \sur{Friend}}\email{jamesfriend@wustl.edu}

\affil[1]{\orgdiv{Medically Advanced Devices Laboratory, Center for Medical Devices, Department of Mechanical and Aerospace Engineering}, \orgname{University of California San Diego}, \orgaddress{\city{La Jolla}, \state{CA}, \postcode{92093-0411}, \country{USA}}}

\affil[2]{\orgdiv{Propulsion and Energy Research Lab, Department of Mechanical and Aerospace Engineering}, \orgname{University of Central Florida}, \orgaddress{\city{Orlando}, \state{FL}, \postcode{32816}, \country{USA}}}

\affil[3]{\orgdiv{Medically Advanced Devices Laboratory, Department of Mechanical Engineering and Materials Science}, \orgname{Washington University in St.\ Louis}, \orgaddress{\street{1 Brookings Drive}, \city{St.\ Louis}, \state{MO}, \postcode{63130}, \country{USA}}}

\abstract{This experimental study demonstrates MHz-order acoustic control of Rayleigh-regime liquid-jet breakup using a compact through-hole lithium niobate thickness-mode transducer integrated at the nozzle exit and driven near its \SI{6.7}{MHz} resonance. Jet dynamics were measured by high-speed shadowgraphy at \SI{50000}{fps}, and breakup lengths were obtained from instantaneous jet-length records using an autocorrelation-based method that identifies statistically separated breakup events. Acoustic actuation produced visible interfacial corrugations and reduced the normalized breakup length, $L_b/d$, for both water and glycerol--water jets. In water, the response was threshold-like at low Weber number and weakened as the jet velocity increased to $\mathrm{We}\simeq 50$. A reduced-order threshold correlation, fit to the water amplitude and velocity sweeps, captured this abrupt shortening and the subsequent loss of actuation efficacy. More viscous glycerol--water mixtures shortened immediately at low forcing, indicating a distinct acoustic-coupling regime rather than a universal collapse of the water response. A wetting-control experiment showed that liquid spreading at the nozzle base can also shorten the jet, so the measured response reflects coupled radiation-pressure forcing, capillary-wave feedback, and wetting-mediated disturbance injection. Resolved droplet tracking further showed that MHz excitation changes the visible gate-crossing droplet population, although sub-resolution mist and micron-scale atomization were not measured. These results establish through-hole MHz actuation as a compact route for modifying free-surface jet breakup and provide a repeatable image-analysis protocol for experimental-fluid-mechanics measurements.}


\maketitle

\section{Introduction}
The breakup of a round liquid jet in quiescent air is governed by the Rayleigh--Plateau instability, in which surface tension causes the jet to fragment into droplets to minimize interfacial area \citep{Rayleigh1878,Eggers2008}. This phenomenon underpins a broad range of industrial and scientific applications, including fuel atomization, inkjet printing, pharmaceutical spray generation, and droplet-based microfluidics \citep{Lin2003,Lefebvre2017}. A persistent practical objective across these fields is the controlled reduction of the breakup length, which can yield earlier droplet production, finer sprays, narrower droplet-size distributions, and more efficient atomization \citep{Lefebvre2017}. The theoretical framework for predicting the natural breakup of a cylindrical jet was established by Rayleigh and extended to include viscous and aerodynamic effects by Weber and later reviews of jet and spray formation \citep{Rayleigh1878,Weber1931,Lin1998,Eggers2008}.

One established route to shortening a jet's breakup length is the controlled introduction of external perturbations. Mechanical excitation via piezoelectric elements driving pressure fluctuations inside a liquid chamber has been studied extensively in the context of continuous inkjet printing and forced capillary jets \citep{Donnelly1966,Goedde1970,Kalaaji2003,Lad2016,Yang2019}. Electric field forcing can also destabilize jets \citep{Collins2007}, though this approach is restricted to conducting or dielectric fluids. Magnetic forcing requires magnetically susceptible media \citep{Shimasaki2011}. Each of these methods imposes constraints on the fluid properties, device geometry, or forcing pathway.

Acoustic radiation pressure offers a comparatively fluid-agnostic route, since it acts on liquid--gas interfaces regardless of the liquid's electrical or magnetic properties. However, many acoustic or vibration-based jet-breakup studies have employed frequencies below the MHz range, where the dominant mechanism is direct mechanical modulation of the nozzle, chamber, or imposed inlet disturbance \citep{Lad2016,Yang2019}. At those frequencies, the acoustic wavelength in the liquid is much larger than the jet diameter, and the jet responds primarily through hydrodynamic forcing. Acoustic radiation pressure does not necessarily promote capillary breakup. It has previously been used to suppress the Rayleigh–Plateau instability of liquid bridges through externally applied radiation stresses \citep{MarrLyon1997,MarrLyon2001}. More directly, \citet{Bertin2010} demonstrated that an ultrasonic field injected axially and guided along a liquid column can generate radial radiation stresses that stabilize the column against capillary disturbances.

The physics qualitatively changes at MHz-order frequencies. \citet{Frommhold:2014aa} used a 1.8\,MHz disturbance to create monodisperse droplets, though the acoustic waves were not applied at the nozzle exit. For MHz-order acoustic excitation of sessile droplets, visible capillary waves can arise through a pressure--interface feedback mechanism after a finite onset threshold \citep{Zhang2023}. That observation motivates the question of whether analogous MHz-order forcing can be coupled into a free cylindrical jet at the nozzle exit to control its breakup length.

The nozzle orifice exit is a natural coupling point for this mechanism. The surface disturbance that ultimately produces pinch-off is convected downstream within the fluid jet from the orifice exit, and the breakup length depends logarithmically on the effective initial disturbance amplitude. An acoustic source placed directly at the exit can therefore modify both the initial capillary disturbance and the local wetting state. In the present system, a lithium niobate thickness-mode transducer is drilled to form the flow passage itself, so the liquid exits through a vibrating piezoelectric element rather than through a passive nozzle followed by a remote forcing region.

In this work, we demonstrate that MHz-order acoustic excitation oriented along the jet propagation direction and applied directly at the nozzle exit produces a pronounced reduction in the Rayleigh-regime breakup length of cylindrical liquid jets in quiescent air. Driven at 6.7~MHz, corresponding to a resonance of the piezoelectric transducer comprising the orifice exit, the excitation frequency is far above the natural Rayleigh-Plateau instability frequency of the fluid jet. Using the notation of \citet{Eggers2008}, using such a high frequency produces a reduced wavenumber $x = \frac{\pi d}{\lambda} = 14.2$, where $d$ is the orifice diameter and $\lambda$ is the axial wavelength of the driving disturbance, far above the range $0<x<1$ over which the jet would be driven unstable via the Rayleigh-Plateau instability.\footnote{The convected form of this reduced wavenumber in \citet{Eggers2008} is likewise very large: $x_\text{conv} = \frac{\pi d f}{v_j} \approx 3.3 \times 10^4$ with $f$ as the excitation frequency and $v_j$ the jet velocity. Either way, the direct excitation of the Rayleigh-Plateau instability from such a high frequency is implausible.} We integrate a through-hole thickness-mode lithium niobate transducer into a compact nozzle assembly driven near a \SI{7}{MHz} resonance and characterize jet dynamics using high-speed shadowgraphy at \SI{50000}{fps}. We introduce an autocorrelation-based image-analysis method to isolate statistically separated breakup lengths $L_b$ from instantaneous jet-length time series. We then show how the normalized breakup length depends on the normalized acoustic forcing amplitude $u^*=\dot{\xi}_p/u_0$, the Weber number $\mathrm{We}$, the Ohnesorge number $\mathrm{Oh}$, and the nozzle-base wetting state. We also quantify optically resolved gate-crossing droplet distributions above a conservative pixel-scale cutoff, allowing the breakup-length measurements to be compared with the size distribution of visible detached fragments. Finally, we include a reduced-order correlation for the water data. Its longer derivation is placed in the Supplementary Information.

\section{Experimental methods}

\begin{figure}[t]
\centering
\maybeincludegraphics[width=0.95\linewidth]{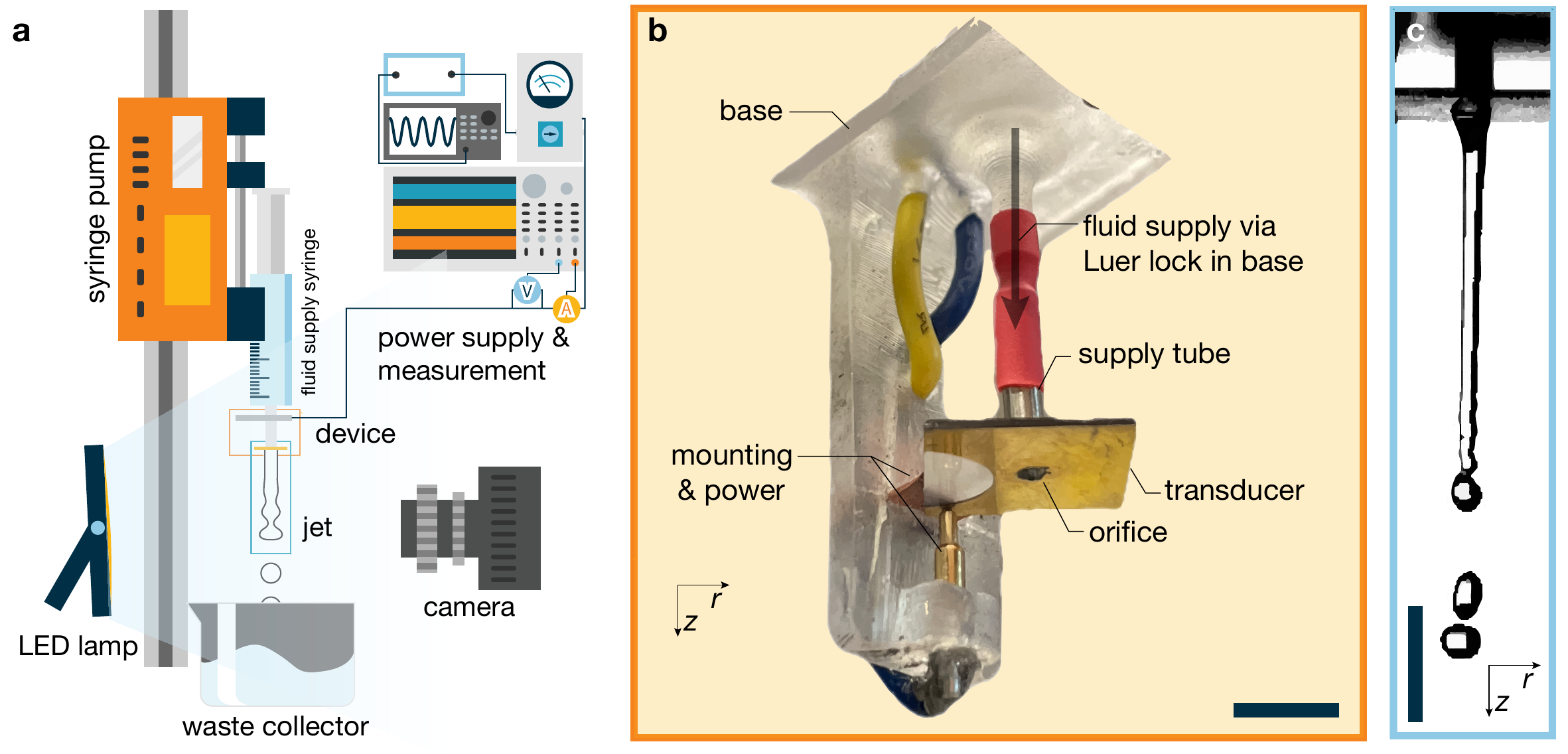}
\caption{Experimental arrangement for acoustic control of liquid-jet breakup. (a) Schematic of the syringe-pump, electrical-actuation, and high-speed-imaging setup. For each condition, 46,560 frames were recorded at 50,000 fps. (b) Close-up of the nozzle coupled to the through-hole thickness-mode transducer and compact electrical connection. (c) Representative shadowgraph image of the cylindrical jet. Both scale bars are 5 mm.}
\label{fig:setup}
\end{figure}

A compact through-hole lithium niobate thickness-mode transducer was used as the nozzle so that acoustic excitation was applied directly at the jet exit. The device was fabricated from a \SI{500}{\micro\meter}-thick, 127.68$^{\circ}$ Y-rotated lithium niobate chip with Ti/Au electrodes on both faces. A \SI{1}{mm} through-hole was drilled through the chip and aligned with a stainless-steel tube inside a three-dimensional printed fixture (Fig.~\ref{fig:setup}). The transducer was driven near its thickness resonance using a signal generator and RF amplifier, while voltage and current were measured at the device to estimate the delivered electrical input. Surface particle velocity $\dot{\xi}_p$ at resonance was measured separately by laser Doppler vibrometry and was used to define the normalized acoustic forcing $u^*=\dot{\xi}_p/u_0$. Full fabrication and characterization details are provided in Supplementary Sec.~S1.%~\ref{sec:supp_methods}.

Flow was supplied by a syringe pump and independently checked gravimetrically. After each change in flow rate or acoustic forcing, the jet was allowed to reach steady operation for \SI{5}{s} before recording. The jet was backlit with a diffuse LED source and imaged at \SI{50000}{fps}. Experiments were conducted with DI water over $Q=\SI{30}{mL\per min}$--\SI{90}{mL\per min} and with glycerol--water mixtures at $Q=\SI{30}{mL\per min}$. All primary parametric sweeps were performed on a single nozzle--transducer assembly to avoid changes in exit geometry. The corresponding flow conditions were expressed as
\begin{equation}
\mathrm{We}=\frac{\rho u_0^2 d}{\sigma}, \qquad
\mathrm{Oh}=\frac{\mu}{\sqrt{\rho\sigma d}}, \qquad
\mathrm{Re}=\frac{\rho u_0 d}{\mu},
\label{eq:dimensionless_groups}
\end{equation}
where $u_0$ is the mean exit velocity, $d=\SI{1}{mm}$ is the nozzle diameter, and $\rho$, $\sigma$, and $\mu$ are the liquid density, surface tension, and dynamic viscosity. Fluid properties, flow conditions, and processing thresholds are listed in Supplementary Tables~S1 and~S2. %~\ref{tab:s_fluid_properties} and \ref{tab:s_droplet_threshold_settings}. 
All data associated with this manuscript is available online at \href{https://doi.org/10.6075/J02N52F5}{https://doi.org/10.6075/J02N52F5}.

\begin{figure}[t]
\centering
\maybeincludegraphics[width=0.72\linewidth]{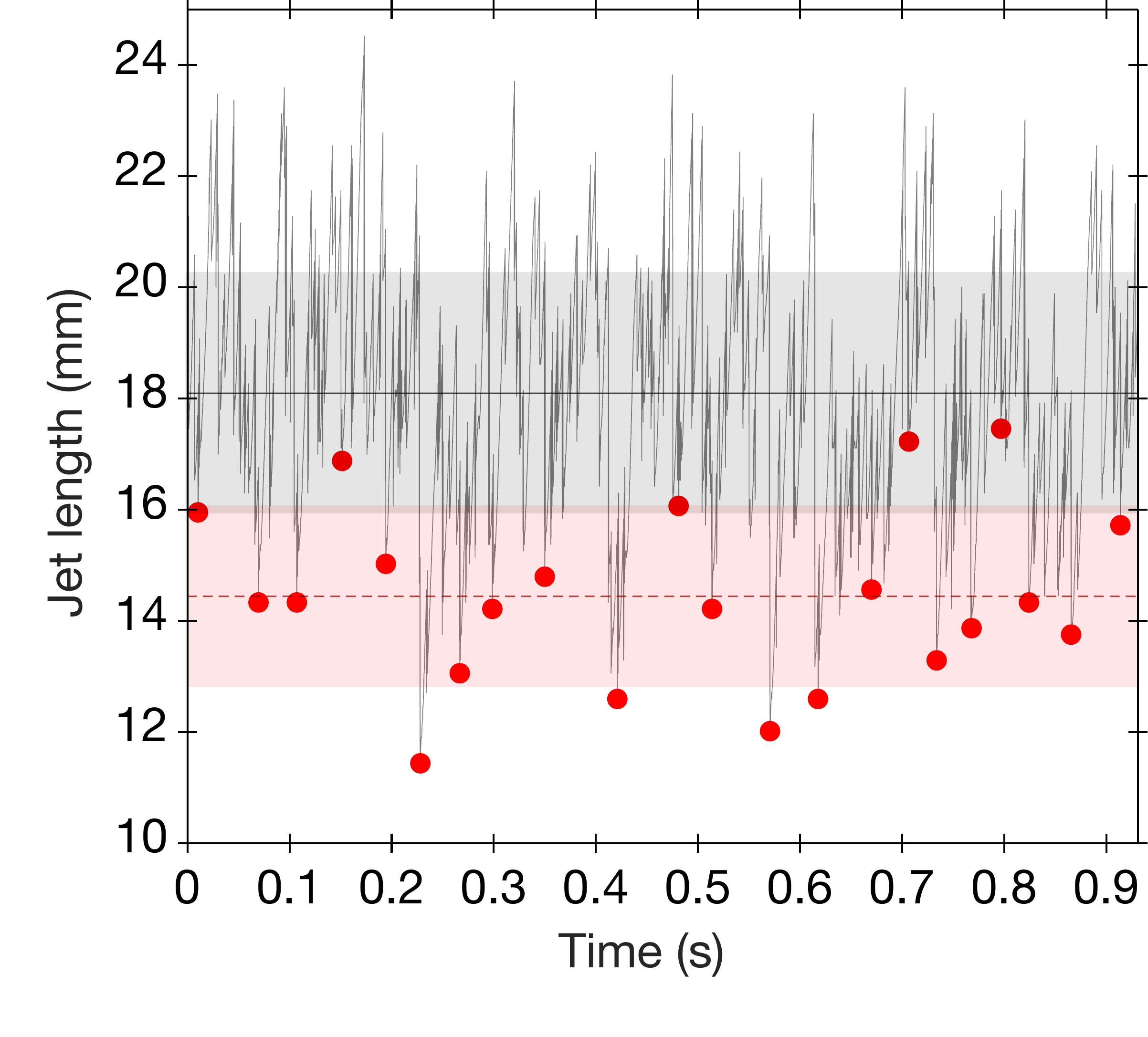}
\caption{Example temporal evolution of the instantaneous jet length $L_j(t)$. The autocorrelation-based event-selection method identifies statistically separated breakup-event minima, marked by red symbols. For this data set, the event-averaged breakup length is approximately \SI{14}{mm}, whereas directly averaging all correlated frames gives approximately \SI{18}{mm}. Shaded regions indicate one standard deviation about each mean.}
\label{fig:event_selection}
\end{figure}

Breakup length was obtained from high-speed image sequences as an event-averaged intact length. For each frame, the instantaneous jet length $L_j(t)$ was defined as the distance from the nozzle exit to the first liquid disconnection. Candidate breakup events were identified as local minima of $L_j(t)$, retained only when their prominence exceeded 30\% of the video-specific jet-length range and when adjacent retained minima were separated by at least twice the first-zero autocorrelation time of $L_j(t)$. The reported breakup length $L_b$ is the mean of these statistically separated event minima and is reported as $L_b/d$; error bars denote the standard deviation of the retained event-wise lengths. Comparison between these two approaches is shown in Fig.~\ref{fig:event_selection}.

Resolved detached droplets were analyzed from shadowgraph videos using background-difference segmentation, contour filling, nearest-neighbor tracking, and a virtual downstream gate. Each accepted droplet was counted once when its centroid crossed the gate in the flow direction. The equivalent projected diameter was calculated as $D_{\mathrm{eq}}=\sqrt{4A/\pi}\,s$, where $A$ is the filled contour area and $s=\SI{0.11594}{mm\per pixel}$ is the image calibration. Quantitative droplet-size statistics were restricted to $D_{\mathrm{eq}}\ge 3s\simeq\SI{0.35}{mm}$; smaller droplets and any micron-scale mist were treated as unresolved. Further details are described in Supplementary Sec.~S1%~\ref{sec:supp_methods}.

\section{Results and discussion}
The results are organized around the physical requirements for MHz-order acoustic control of a Rayleigh-regime jet. First, we verify that the acoustic field can be guided by the liquid jet rather than being confined to the nozzle exit. We then examine how the breakup length changes with acoustic forcing amplitude and jet velocity for water. Because nozzle-base wetting can alter the effective exit condition, we next isolate its contribution before correlating the water amplitude and velocity sweeps with a reduced-order threshold correlation. We then examine fluid viscosity and the distribution of resolved gate-crossing droplets.

\subsection{Acoustic transmission along the jet}
A necessary condition for acoustic control is that the pressure field launched at the nozzle exit can reach the region where capillary breakup occurs. A liquid jet surrounded by air can act as an acoustic waveguide because the liquid--air interface provides a strong acoustic impedance contrast. \citet{Lonzaga2007} showed experimentally that internally excited ultrasonic radiation pressure can be guided by a free liquid jet and that the relevant cutoff frequency scales inversely with the jet diameter. For the lowest axisymmetric pressure-release mode, the cutoff frequency may be estimated as
\begin{equation}
f_{\mathrm{co}}=\frac{c}{2\pi}\frac{2.4048}{a},
\label{eq:cutoff}
\end{equation}
where $c$ is the sound speed in the liquid and $a$ is the local jet radius. Propagation of this mode requires the driving frequency to exceed the local cutoff frequency; below cutoff, the corresponding guided field is evanescent.

For water, using $c=\SI{1480}{m\per s}$ and $f=\SI{6.7}{MHz}$ gives a limiting radius of $a\approx\SI{80}{\micro\meter}$. The intact jets observed here have local radii of order 100--\SI{500}{\micro\meter} before breakup, so the \SI{6.7}{MHz} excitation is above the estimated cutoff along the observed intact length. This supports the interpretation that acoustic radiation pressure can act as a distributed forcing along the jet, rather than only as a localized mechanical disturbance at the nozzle. The same conclusion is consistent with the visual appearance of surface corrugations over the intact portion of the acoustically actuated jet.

\subsection{Effect of acoustic forcing amplitude}
In the absence of ultrasound, the jet breaks up into droplets via the Rayleigh--Plateau instability. For an inviscid cylindrical jet, the most amplified varicose mode satisfies $kR\simeq 0.697$, corresponding to
\begin{equation}
\lambda_{\max}\simeq 4.51d,
\label{eq:rayleigh_wavelength}
\end{equation}
where $R=d/2$ is the jet radius \citep{Rayleigh1878,Eggers2008}. This wavelength is the dominant disturbance wavelength or droplet-spacing scale, not the breakup length itself. The measured breakup length is instead a convective growth distance over which an initial disturbance grows from an effective amplitude $a_0$ to a nonlinear pinch-off amplitude. In the simplest convective picture,
\begin{equation}
L_b\sim \frac{u_0}{s_{\max}}\ln\left(\frac{R}{a_0}\right),
\label{eq:convective_growth_intro}
\end{equation}
where $u_0$ is the jet velocity and $s_{\max}$ is the temporal growth rate of the dominant unstable mode \citep{Lin1998,Eggers2008}. Thus, $L_b$ can span several Rayleigh wavelengths and is sensitive to the initial perturbation field, nozzle geometry, wetting state, and residual mechanical vibration.

\begin{figure}[t]
\centering
\maybeincludegraphics[width=0.75\linewidth]{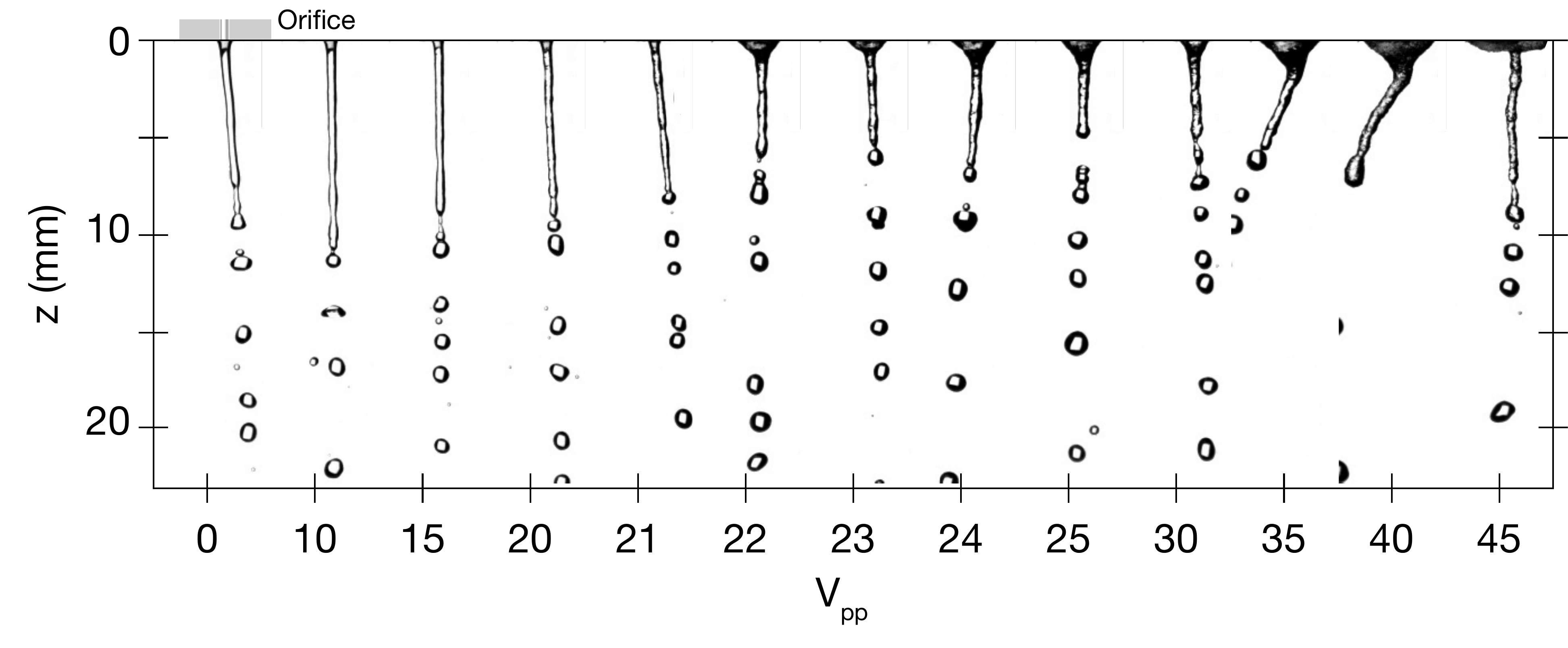}
\caption{Representative snapshots of the water jet at increasing input voltages with $\mathrm{We}=5.5$ and $\mathrm{Oh}=0.0038$. The horizontal axis is the input voltage measured in peak-to-peak volts and is not linearly scaled.}
\label{fig:voltage_snapshots}
\end{figure}

\begin{figure}[t]
\centering
\maybeincludegraphics[width=0.68\linewidth]{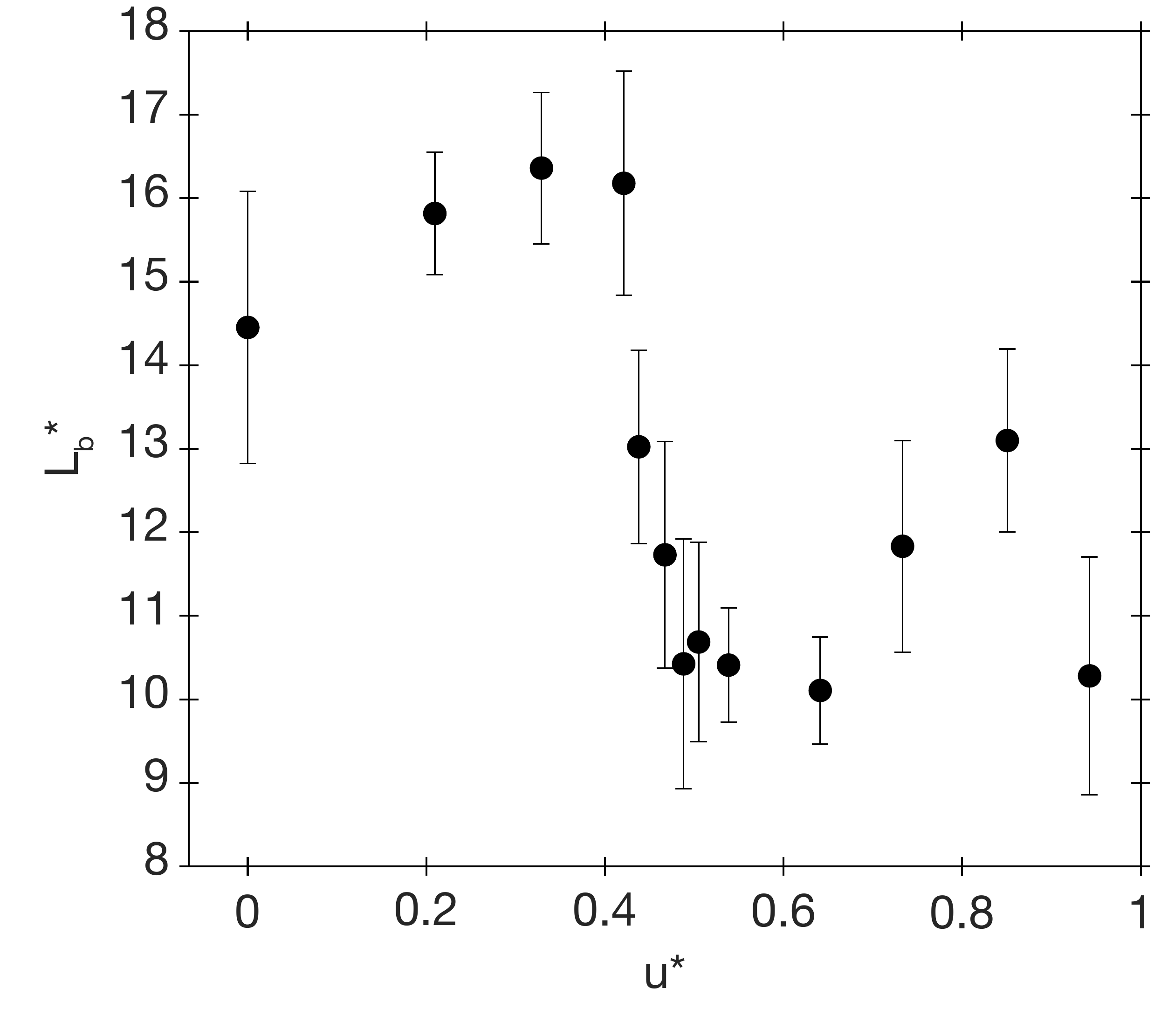}
\caption{Average normalized breakup length $L_b^*=L_b/d$ as a function of normalized acoustic velocity $u^*=\dot{\xi}_p/u_0$ for water at $\mathrm{Re}=611$, $\mathrm{We}=5.5$, and $\mathrm{Oh}=0.0038$. Each point represents the average of five or more physically independent recordings; error bars denote the standard deviation of retained breakup-event lengths.}
\label{fig:water_amplitude}
\end{figure}

The acoustic response is reported using the normalized breakup length $L_b^*=L_b/d$ and normalized acoustic velocity $u^*=\dot{\xi}_p/u_0$, where $\dot{\xi}_p$ is the transducer particle velocity measured by laser Doppler vibrometry at resonance. Figure~\ref{fig:voltage_snapshots} shows that acoustic actuation qualitatively changes the breakup morphology. Without ultrasound, the jet exhibits familiar varicose growth of capillary disturbances and breaks after several jet diameters. With increasing acoustic drive, interfacial corrugations become visible on the intact jet and the first disconnection shifts upstream. The corresponding statistics in Fig.~\ref{fig:water_amplitude} show a nonmonotonic response. At small $u^*$, $L_b^*$ changes weakly and can increase slightly. Once $u^*$ exceeds a threshold range, $L_b^*$ decreases abruptly. At still larger $u^*$, the reduction saturates and partially recovers.

This behavior differs from the simplest small-amplitude forced-jet picture. In classical forced capillary jets, increasing the imposed disturbance amplitude generally shortens the breakup length because the logarithmic factor in Eq.~\eqref{eq:convective_growth_intro} decreases \citep{Kalaaji2003}. Audio-frequency experiments showed that controlled low-frequency perturbations can select the breakup wavelength and regularize droplet formation \citep{Donnelly1966,Goedde1970}. In contrast, the present excitation frequency is MHz-order, far above the hydrodynamic droplet-formation frequency. The interface therefore does not respond as though it were being directly shaken at the breakup frequency. Instead, the response is more consistent with acoustic radiation pressure driving a lower-frequency capillary-wave envelope through a pressure--interface feedback mechanism. A similar separation of time scales was reported by \citet{Zhang2023}, who showed that MHz acoustic excitation of a sessile droplet can generate visible low-frequency capillary waves only after a finite threshold is reached.

The partial recovery of $L_b^*$ at the largest forcing amplitudes should be interpreted cautiously. Large-amplitude forced jets can exhibit breakup reversal, in which the breakup length stagnates or increases as forcing is increased further. \citet{Denner2022} attributed this reversal to nonlinear internal flow structures, including vortex rings and local flow obstruction, that can slow or reverse neck collapse. The present system differs because the forcing is ultrasonic and is coupled through the nozzle base rather than imposed as a prescribed inlet-velocity modulation. Nevertheless, the analogy is useful: after the acoustic field has triggered upstream capillary breakup, additional input energy may be diverted into base wetting, acoustic streaming, internal recirculation, satellite-drop formation, or competing interfacial modes rather than monotonically amplifying the same destabilizing Rayleigh mode.

\begin{figure}[t]
\centering
\maybeincludegraphics[width=0.88\linewidth]{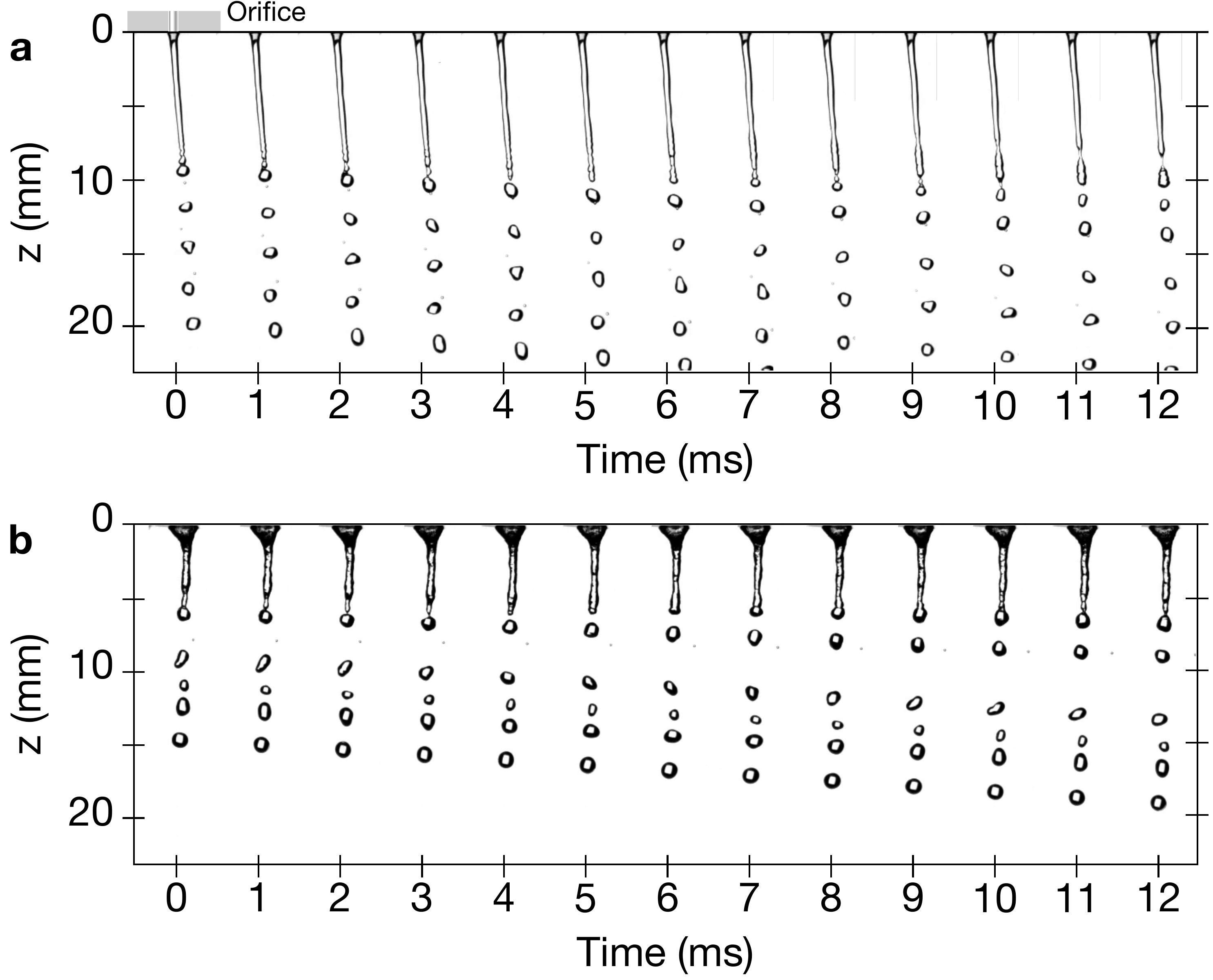}
\caption{Representative breakup sequences for water with (a) no ultrasound and (b) sinusoidal acoustic excitation at 25 Vpp. Conditions are $\mathrm{Re}=611$, $\mathrm{We}=5.5$, and $\mathrm{Oh}=0.0038$. Acoustic excitation shifts the first disconnection upstream and produces a more corrugated intact jet.}
\label{fig:breakup_sequences}
\end{figure}

\subsection{Effect of jet velocity}
\begin{figure}[t]
\centering
\maybeincludegraphics[width=0.6\linewidth]{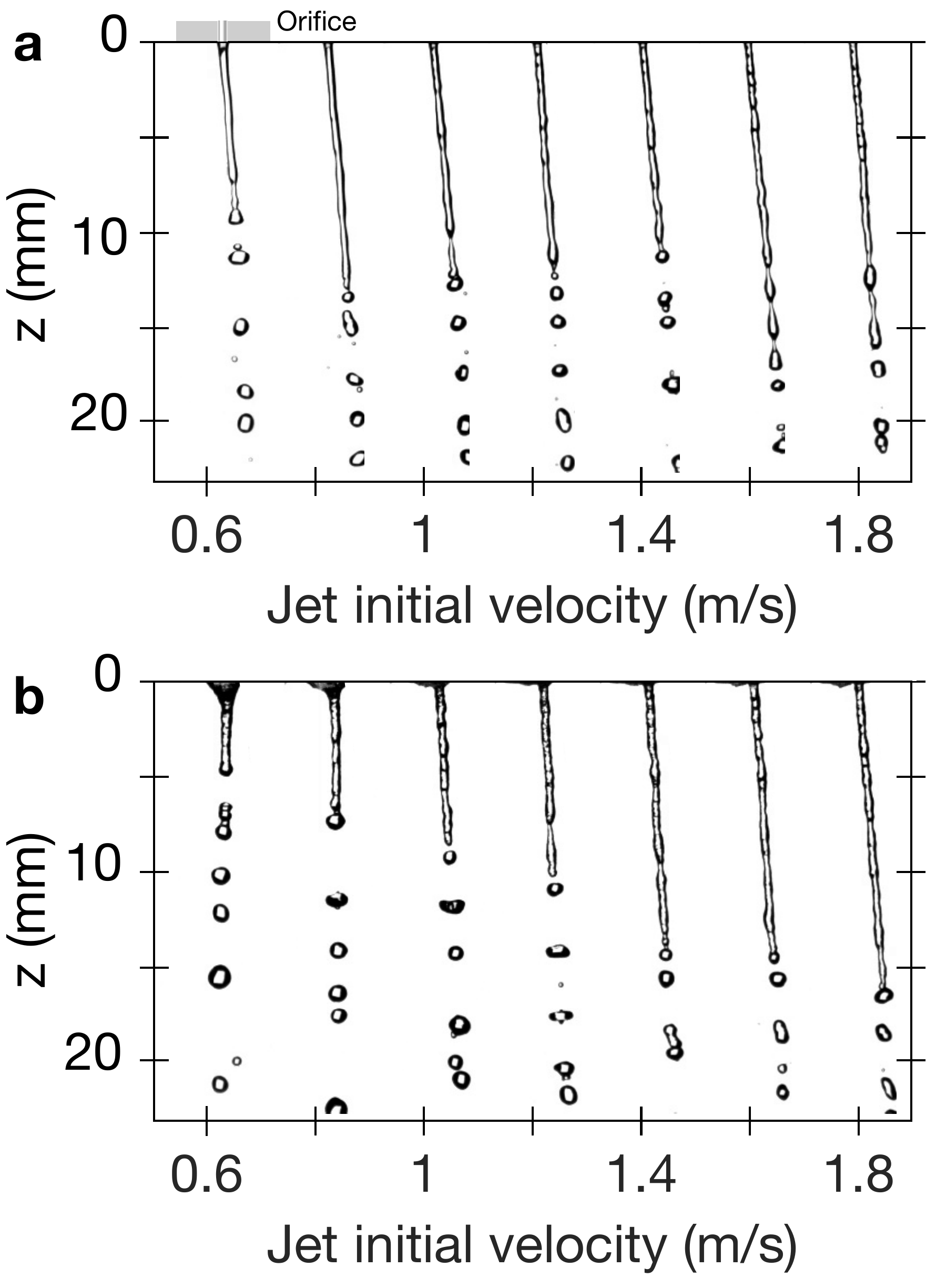}
\caption{Representative breakup sequences at different initial jet velocities for water with (a) no ultrasound and (b) 25 Vpp acoustic excitation. The corresponding $\mathrm{Re}$ and $\mathrm{We}$ values are listed in Supplementary Table~S1.}%~\ref{tab:s_fluid_properties}.}
\label{fig:velocity_snapshots}
\end{figure}

\begin{figure}[t]
\centering
\maybeincludegraphics[width=0.74\linewidth]{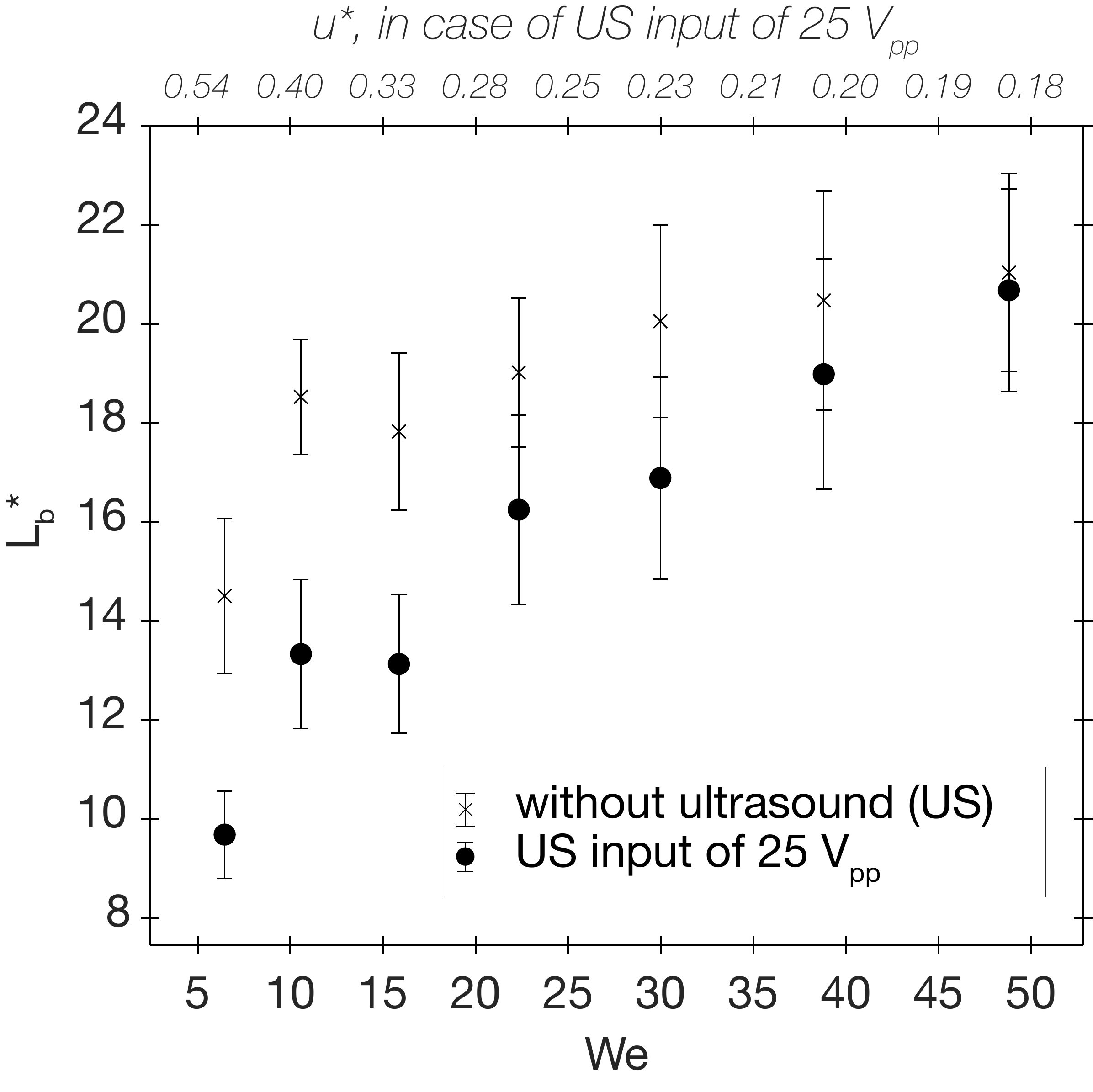}
\caption{Average normalized breakup length as a function of $\mathrm{We}$ for water with and without 25 Vpp ultrasound. The upper axis shows the corresponding $u^*=\dot{\xi}_p/u_0$ for the acoustic case and is included only for comparison.}
\label{fig:water_velocity}
\end{figure}

Figures~\ref{fig:velocity_snapshots} and \ref{fig:water_velocity} show that acoustic actuation is most effective at low jet velocity. In the unforced case, $L_b^*$ generally increases as $\mathrm{We}$ increases, consistent with a convective instability whose capillary growth rate does not increase as rapidly as the downstream transport speed. When the same 25 Vpp acoustic input is applied, the breakup length is substantially reduced at low $\mathrm{We}$, but the difference between the forced and unforced cases decreases as the jet flow increases, reducing the residence time of the jet, and reducing the overall influence of the oscillation in the jet. At the largest flow rates, the two curves approach one another.

Two effects likely contribute to this trend. First, at fixed voltage the measured transducer velocity is approximately fixed while $u_0$ increases, so the normalized forcing $u^*=\dot{\xi}_p/u_0$ decreases. The acoustic forcing therefore becomes weaker relative to jet inertia. Second, the residence time over a given downstream distance decreases as $u_0$ increases, leaving less time for the acoustically seeded disturbance to grow before it is convected away. This is analogous to conventional forced-jet experiments, where the breakup response depends not only on disturbance amplitude but also on the matching between imposed disturbance and hydrodynamic growth process \citep{Donnelly1966,Kalaaji2003}.

\subsection{Effect of base wetting}
\label{sec:wetting}
In addition to direct acoustic forcing, wetting at the nozzle exit influences the measured breakup length. Apparent from Figs.~\ref{fig:voltage_snapshots} and~\ref{fig:breakup_sequences}, the fluid wets the orifice exit when applying ultrasound. This wetting widens the jet at the orifice exit to a diameter considerably larger than the orifice and the jet farther downstream. High-frequency acoustic waves propagating within a solid substrate can alter dynamic wetting and contact-line motion through the combined action of acoustic radiation pressure and boundary-layer-driven streaming. Depending on the initial surface wettability and forcing conditions, substrate vibration can promote droplet spreading or draw a thin acoustowetting film from a sessile liquid volume \citep{Rezk:2012fk,Rezk:2014rr}. Because nozzle geometry and inlet disturbances are known to influence jet morphology and breakup behavior \citep{McCarthy1974}, this wetted region must be treated as part of the acoustic coupling pathway rather than as a purely incidental artifact. Indeed, efforts have been made in the past to avoid such wetting, as they can lead to orifice fouling and print failure in ink-jet printers \citep{Borgelink2022,DeJong2007}.

To isolate the wetting contribution, three conditions were compared using DI water at $Q=\SI{30}{mL\per min}$. First, a control jet was recorded under the standard operating condition. Second, the flow rate was temporarily increased to \SI{90}{mL\per min} to remove the wetted base; the flow rate was then returned to \SI{30}{mL\per min} and the jet was recorded after confirming that no visible wetting remained at the nozzle exit. Third, a liquid droplet was intentionally placed at the nozzle exit to create a deliberately wetted base before recording.

\begin{figure}[t]
\centering
\maybeincludegraphics[width=0.58\linewidth]{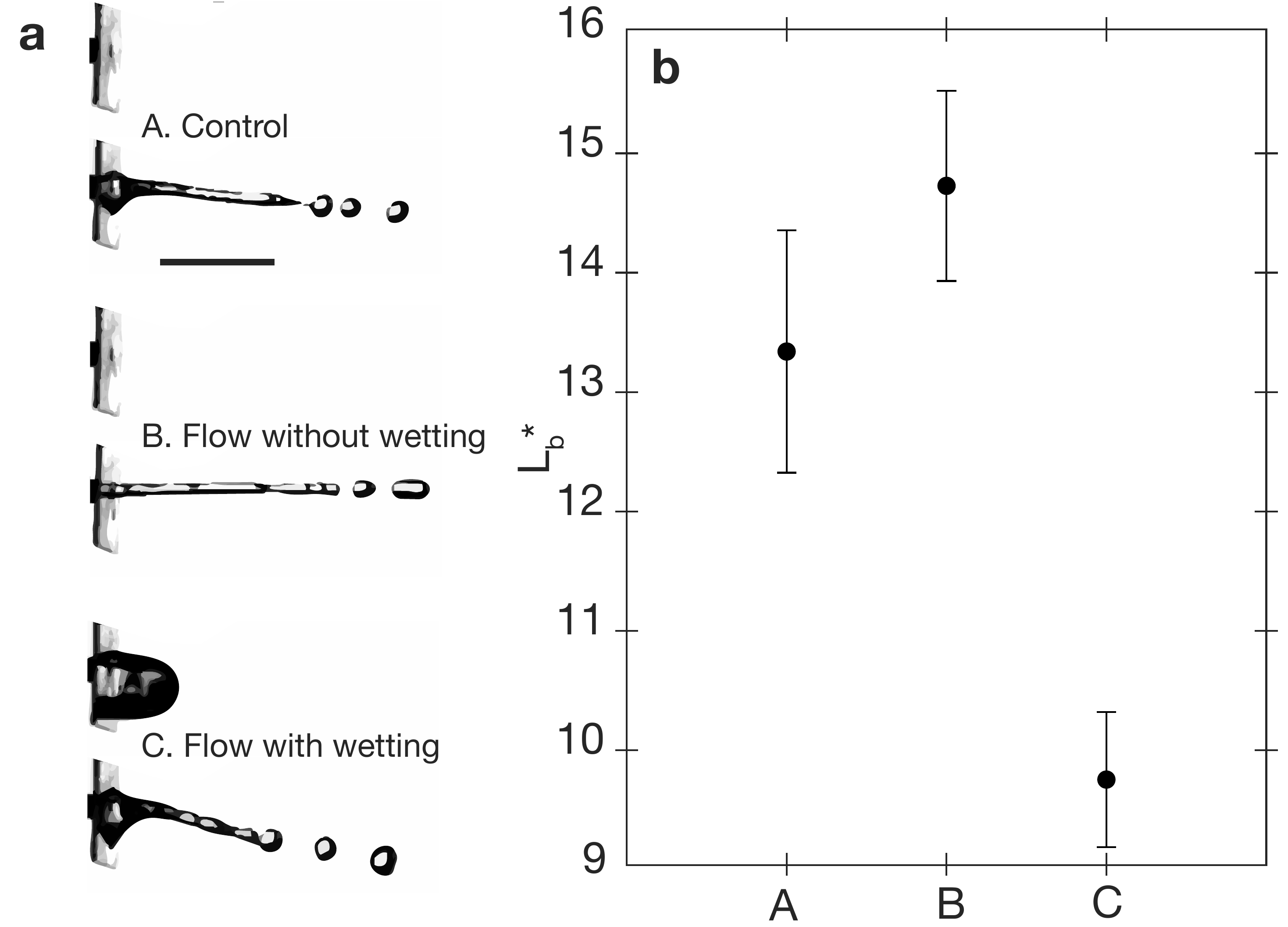}
\caption{Effect of base wetting on jet breakup length. (a) Representative snapshots of DI water jets under three nozzle-base conditions: A, control; B, flow without visible base wetting after high-flow conditioning; and C, flow with an intentionally introduced wetted base. (b) Normalized breakup length $L_b^*$ for the three conditions. Removing the wetted base increases $L_b^*$, whereas introducing a wetted base substantially reduces $L_b^*$, indicating that base wetting contributes to upstream breakup. Error bars denote the standard deviation of retained breakup-event lengths.}
\label{fig:wetting}
\end{figure}

As shown in Fig.~\ref{fig:wetting}, removing the wetted base increases $L_b^*$ relative to the control, whereas intentionally introducing a wetted base substantially shortens the jet. This confirms that base wetting alone can reduce the breakup length. Mechanically, the wetted base can alter the effective exit shape, provide a larger oscillating liquid--solid contact area, and convert the transducer's thickness-mode motion into pressure disturbances normal to the liquid--air interface. The wet base therefore acts as both a geometric perturbation and an acoustic coupling element. The total breakup-length reduction observed under ultrasound should consequently be interpreted as a combined effect of guided acoustic radiation pressure, capillary-wave feedback, and wetting-mediated disturbance injection at the nozzle base. At the same time, excessive base wetting may reduce actuation efficiency because the additional liquid volume can absorb and dissipate acoustic energy while also promoting unwanted thin-film atomization on the device surface adjacent the orifice \citep{Collins2012,Huang2019}.

\subsection{Reduced-order correlation of the sudden decrease in water jet breakup}
\label{sec:reduced_correlation}
The water data in Figs.~\ref{fig:water_amplitude} and \ref{fig:water_velocity} show two linked features: the breakup length does not decrease continuously from zero forcing, but instead decreases abruptly after a finite acoustic drive, and the shortening weakens as $\mathrm{We}$ approaches the upper end of the Rayleigh regime. To connect these observations, we constructed a reduced-order threshold correlation for the water jet. The correlation is not intended as a universal prediction for all tested liquids; rather, it provides a compact representation of the water-jet data and places the observed loss of acoustic efficacy near $\mathrm{We}\simeq 50$ on the same footing as the threshold observed in the amplitude sweep. A longer derivation is given in Supplementary Information, Sec.~S2.%\ref{sec:supp_correlation_derivation}.

The underlying hydrodynamic picture is the convective growth of the Rayleigh--Plateau mode. In the absence of acoustic actuation, the measured breakup length may be written in the usual logarithmic form $L_b\sim (u_0/s_{\max})\ln(R/a_n)$, where $R=d/2$, $s_{\max}$ is the growth rate of the dominant capillary mode, and $a_n$ is an effective background disturbance amplitude set by the nozzle, wetting state, and residual mechanical perturbations. Acoustic actuation is represented by an additional capillary-wave amplitude that turns on only when the guided acoustic drive reaching the wetted nozzle base exceeds a threshold. The reduced control parameter is
\begin{equation}
\Theta =\left(\frac{\dot{\xi}_p}{\dot{\xi}_c}\right)^2\frac{1}{1+\mathrm{We}/\mathrm{We}_t},
\label{eq:theta_main}
\end{equation}
where $\dot{\xi}_p$ is the measured transducer particle velocity, $\dot{\xi}_c$ is the critical particle velocity for onset, and $\mathrm{We}_t$ describes the loss of acoustic coupling as increasing inertia reduces the stable wetted base at the nozzle.

The acoustically sustained capillary-wave amplitude is represented as a saturated threshold branch,
\begin{equation}
\hat{A}_c=A_{\max}\left[\frac{\Theta-1}{\Theta_s+\Theta-1}\right]^{1/2}H(\Theta-1),
\label{eq:ac_branch_main}
\end{equation}
where $\hat{A}_c=A_c/R$, $A_{\max}$ is the normalized saturated capillary-wave amplitude, $\Theta_s$ sets the width of the post-onset transition, and $H$ is the Heaviside distribution \citep{Skinner1989}. Substituting this acoustically generated disturbance into the convective-growth expression gives
\begin{equation}
\frac{L_b}{d}=\frac{u_0}{d s_{\max}}\ln\left[\frac{R}{a_n(u_0)+R\hat{A}_c}\right].
\label{eq:lb_correlation_main}
\end{equation}
Here $a_n(u_0)$ is obtained from the no-ultrasound baseline for the corresponding water condition, so device-specific differences in the background nozzle state are not forced into the acoustic parameters.

A joint fit to the water amplitude sweep and water velocity sweep gives $\dot{\xi}_c=\SI{0.29}{m\per s}$, $\mathrm{We}_t=126$, $A_{\max}=0.011$, and $\Theta_s=0.031$. The corresponding saturated acoustic capillary-wave amplitude is $R A_{\max}\approx \SI{5.5}{\micro\meter}$. The fit gives $\chi^2/\mathrm{dof}=1.46$, where dof refers to the number of independent data values remaining after estimating the model parameters as the ``degrees of freedom'' remaining, and the implied cutoff of the acoustic shortening near $\mathrm{We}\approx 50$ is consistent with the observed collapse of the water-jet actuation effect near $\mathrm{We}\approx 49$. Thus, the correlation captures the water experiments: it represents the threshold-like decrease in $L_b/d$, the post-threshold saturation, and the weakening of the effect at high Weber number. Because the correlation was fit to both water sweeps rather than to a single low-Weber-number datum and then tested independently, we do not claim predictive validation.

\begin{figure}[t]
\centering
\maybeincludegraphics[width=0.94\linewidth]{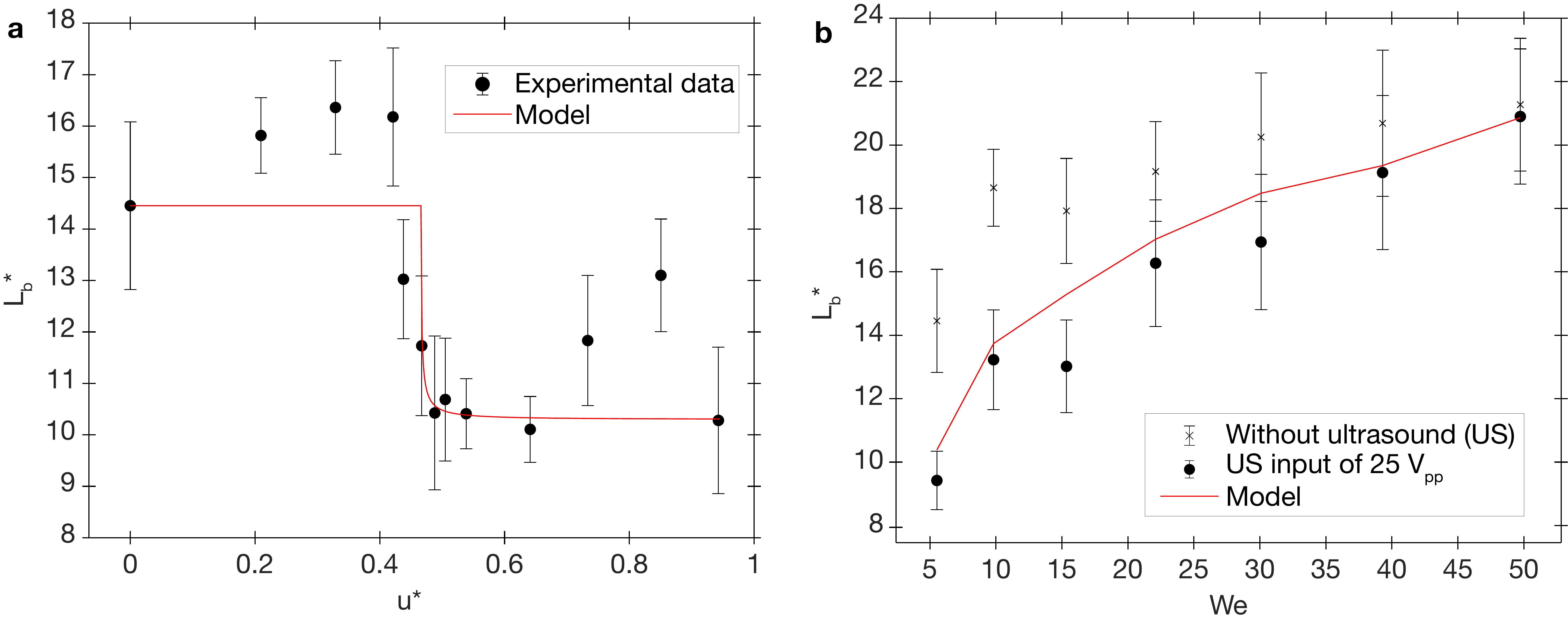}
\caption{Reduced-order threshold correlation for water only. (a) Normalized breakup length as a function of the normalized acoustic velocity $u^*=\dot{\xi}_p/u_0$ for the water amplitude sweep. (b) Normalized breakup length as a function of $\mathrm{We}$ for the water velocity sweep with and without 25 Vpp acoustic excitation. The correlation captures the water data, including the sudden reduction in breakup length and the loss of actuation efficacy near $\mathrm{We}\simeq 50$.}
\label{fig:water_correlation}
\end{figure}

This reduced-order correlation does not extend to the glycerol--water solutions. In the higher-viscosity, higher-Ohnesorge-number fluids, the jet length decreases immediately at low acoustic forcing, without the weak-response plateau and sudden onset observed in water. This immediate response indicates a different acoustic-coupling and breakup regime, likely involving stronger viscous coupling near the wetted base and a different conversion of MHz forcing into the low-frequency interfacial envelope. For this reason, the glycerol--water data are discussed phenomenologically below and are not compared to the threshold breakup uniquely seen with water.

\subsection{Effect of fluid viscosity}
\label{sec:viscosity}
\begin{figure}[t]
\centering
\maybeincludegraphics[width=0.95\linewidth]{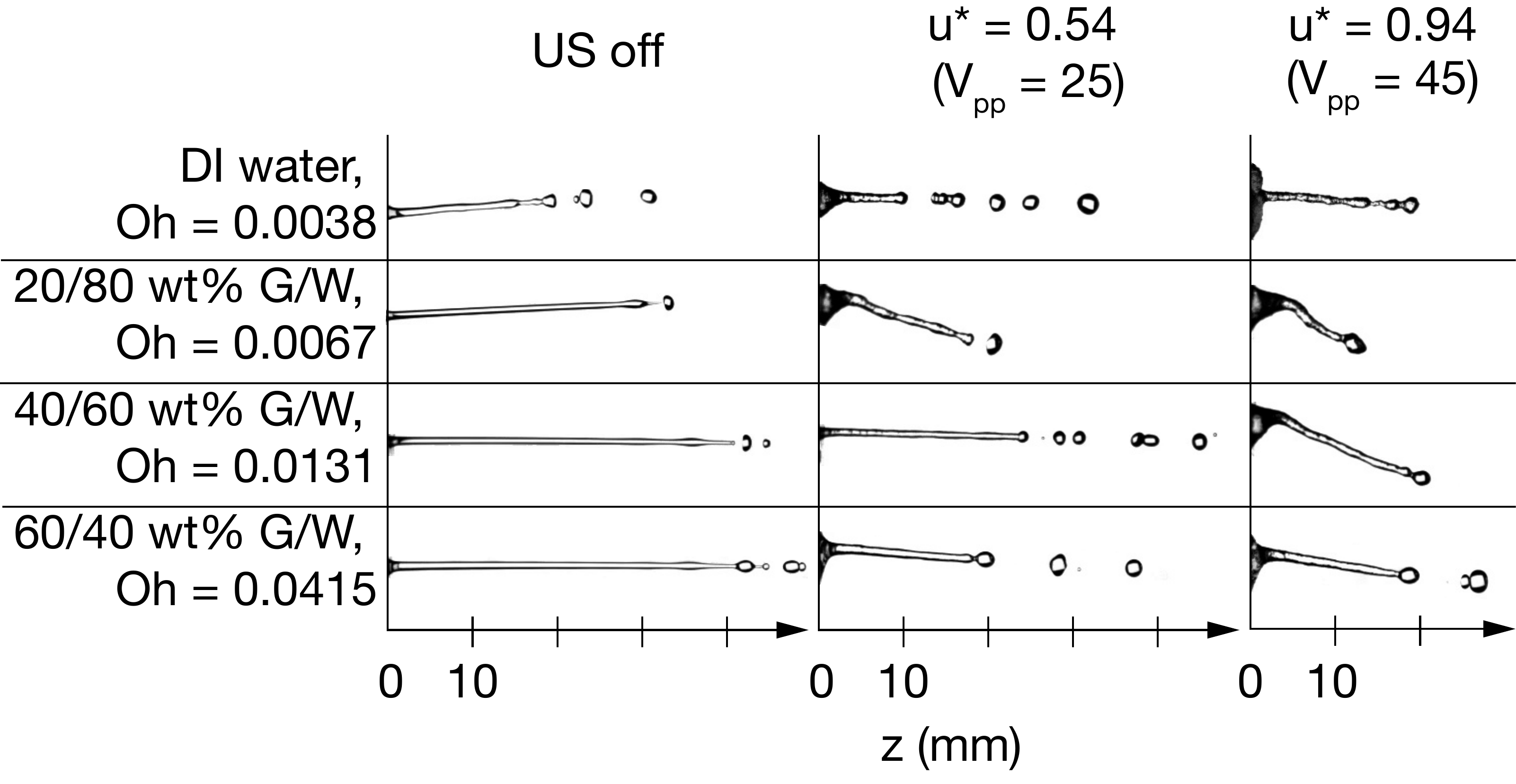}
\caption{Representative snapshots of jet breakup for fluids with different viscosities under acoustic excitation. Rows correspond to DI water, 20/80 wt\% glycerol--water, 40/60 wt\% glycerol--water, and 60/40 wt\% glycerol--water, with $\mathrm{Oh}=0.0038$, 0.0067, 0.0131, and 0.0415, respectively. Columns compare the unforced case, acoustic excitation at 25 Vpp with $u^*=0.54$, and acoustic excitation at 45 Vpp with $u^*=0.94$.}
\label{fig:viscosity_snapshots}
\end{figure}

\begin{figure}[t]
\centering
\maybeincludegraphics[width=0.72\linewidth]{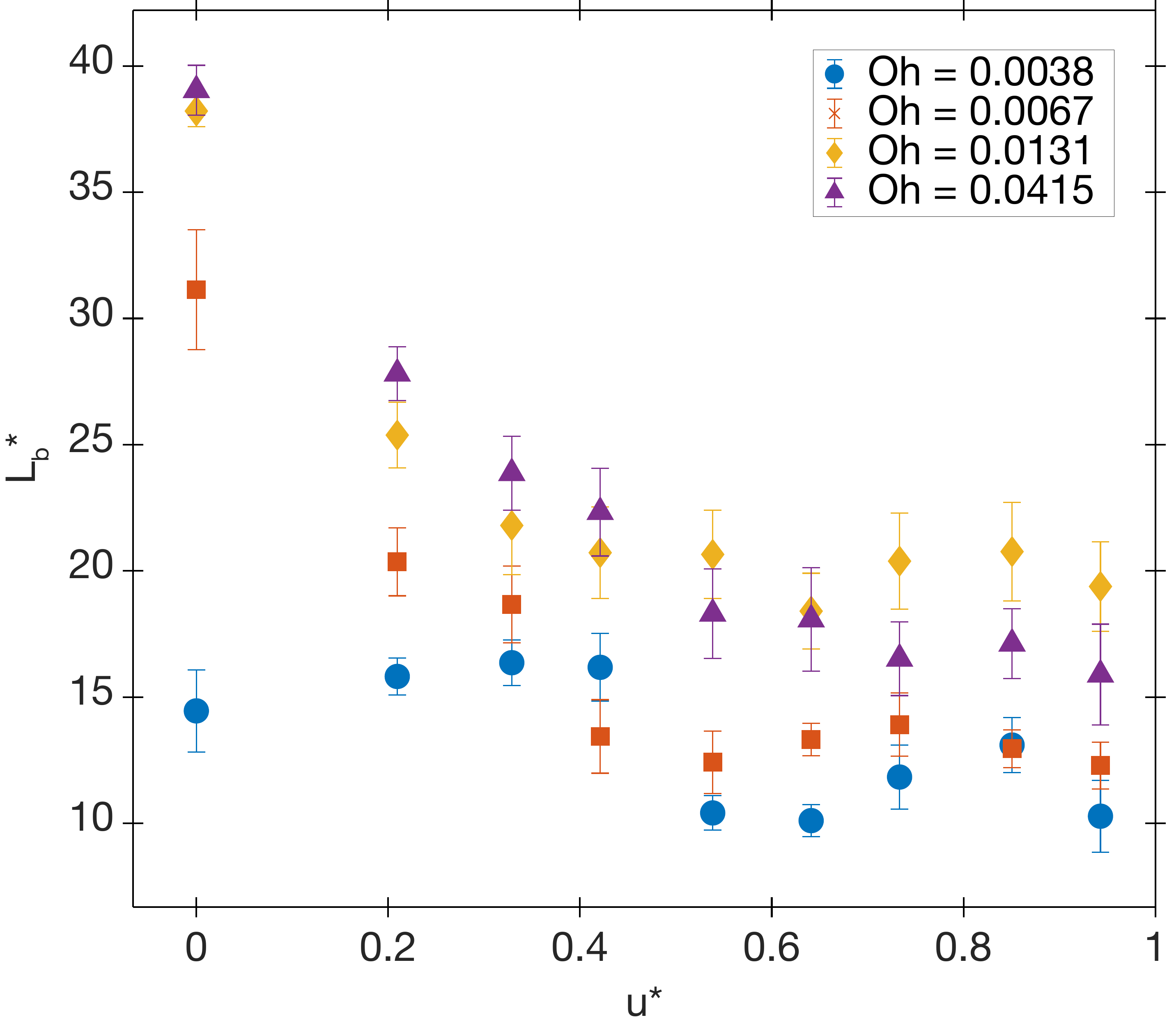}
\caption{Average normalized breakup length for four fluids as a function of normalized acoustic velocity at approximately fixed Weber number, $\mathrm{We}\approx 5.5$--6.8. Increasing $\mathrm{Oh}$ lengthens the unforced jet, while acoustic excitation reduces $L_b^*$ across all tested fluids. The higher-viscosity glycerol--water solutions respond immediately to forcing and are not compared to the water-threshold correlation.}
\label{fig:viscosity_plot}
\end{figure}

The viscosity sweep in Figs.~\ref{fig:viscosity_snapshots} and \ref{fig:viscosity_plot} separates capillary forcing from viscous damping, with weaker effects on surface tension and acoustic impedance through the addition of glycerol. Without ultrasound, increasing $\mathrm{Oh}$ lengthens the breakup length. This agrees with viscous extensions of Rayleigh's theory: viscosity dissipates interfacial motion, reduces the temporal growth rate of unstable capillary modes, and therefore increases the convective distance required for a disturbance to reach pinch-off amplitude \citep{Weber1931,Eggers2008}. Acoustic excitation reduces $L_b^*$ for all four fluids, but the shape of the response changes with $\mathrm{Oh}$.

Water shows a relatively sharp transition from a weak-response branch to a shortened-jet branch. The glycerol--water mixtures exhibit an immediate decrease in $L_b^*$ at low forcing and then remain reduced over a broader range of forcing amplitudes. This contrast indicates that the higher-viscosity, higher-$\mathrm{Oh}$ fluids do not share the same thresholded water response. Viscosity likely plays two opposing roles. On one hand, viscosity damps capillary growth and lengthens the unforced jet. On the other hand, the longer unforced growth distance provides a larger window over which acoustic disturbances can accumulate, and viscous coupling near the wetted base can convert substrate vibration into a lower-frequency interfacial response more directly. Similar sensitivity to $\mathrm{Oh}$ is observed in large-amplitude forced-jet studies, where viscous stresses and finite-amplitude internal flow structures control whether breakup reversal and mode competition appear \citep{Denner2022}.

The present data therefore support a combined mechanism. Acoustic radiation pressure increases the effective initial disturbance amplitude and modifies the pressure distribution near the intact jet, while viscosity controls how this disturbance is dissipated, phase shifted, and converted into hydrodynamic capillary motion. Within the tested range, increased viscosity does not eliminate acoustic control. Instead, it changes the response from the threshold-like behavior observed in water to an immediate shortening regime. At still larger $\mathrm{Oh}$, however, one would expect overdamping to eventually suppress the acoustic effect as the capillary mode itself becomes too strongly damped to grow efficiently.

\subsection{Resolved gate-crossing droplet-size distributions}
\label{sec:droplet_distributions}
\begin{figure}[t]
\centering
\maybeincludegraphics[width=0.65\linewidth]{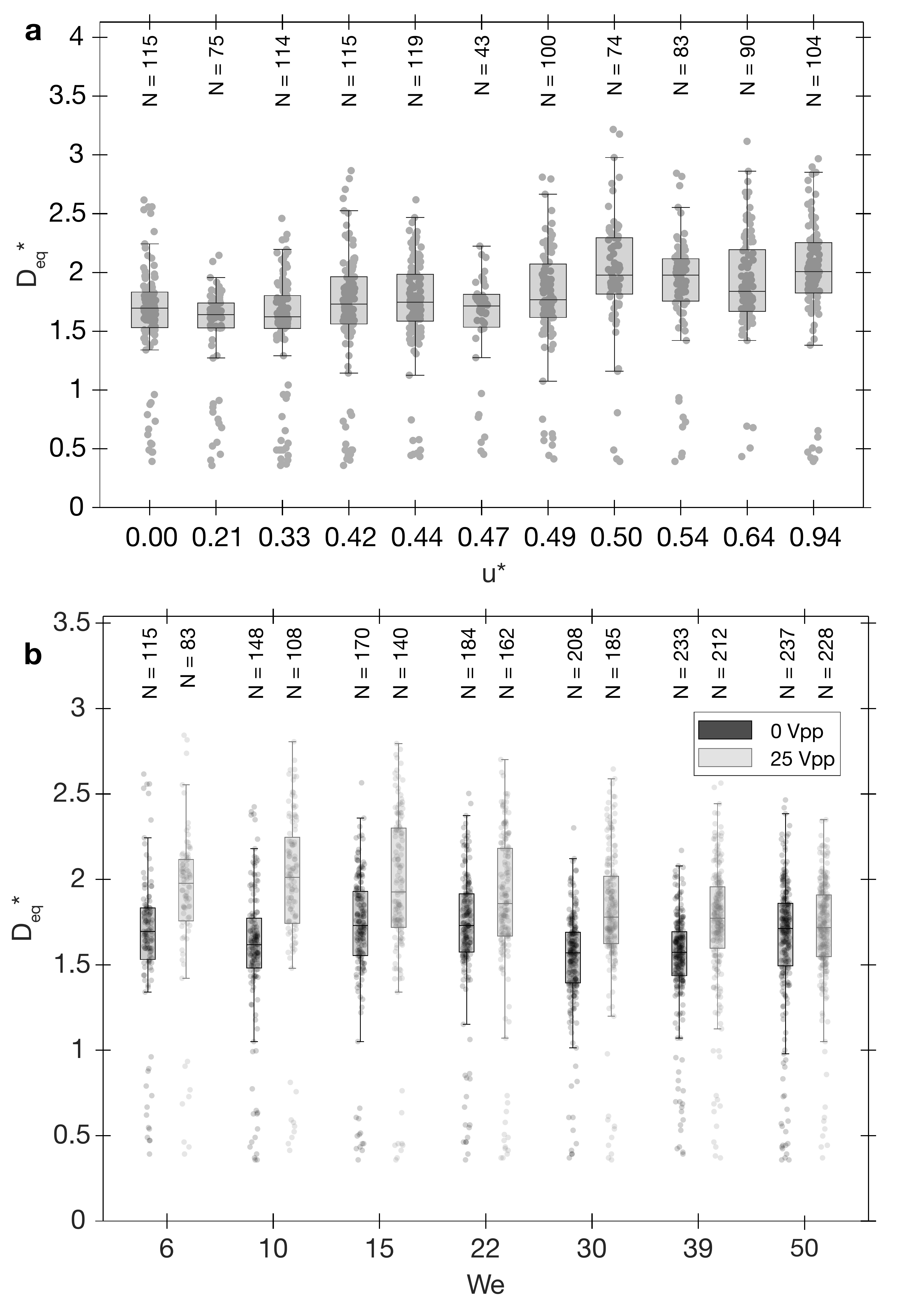}
\caption{Resolved gate-crossing droplet-size distributions. (a) Normalized equivalent droplet diameter $D_{\mathrm{eq}}^*=D_{\mathrm{eq}}/d$ as a function of normalized acoustic velocity $u^*=\dot{\xi}_p/u_0$ for DI water at $Q=\SI{30}{mL\per min}$, $\mathrm{We}=5.56$, and $\mathrm{Oh}=0.0038$. (b) $D_{\mathrm{eq}}^*$ as a function of Weber number for DI water with no ultrasound and with $25~\mathrm{V_{pp}}$ acoustic excitation. Each box shows the median and interquartile range of resolved gate-crossing droplets after the three-pixel cutoff; points are individual retained gate-crossing droplets, and $N$ gives the number of retained droplets in each condition. Since repeat videos of the experiments were not available, significance of differences are not discussed.}
\label{fig:droplet_statistics}
\end{figure}

We next examine the detached visible droplet population. Figure~\ref{fig:droplet_statistics} reports the normalized equivalent diameter $D_{\mathrm{eq}}^*=D_{\mathrm{eq}}/d$ of resolved gate-crossing droplets of size greater than a noise-reducing three-pixel cutoff. For classical Rayleigh breakup of a cylindrical jet, volume conservation applied to the dominant Rayleigh wavelength gives a primary droplet diameter of approximately $D_R\simeq 1.89d$ \citep{Rayleigh1878,Eggers2008}; the resolved gate-crossing droplets in Fig.~\ref{fig:droplet_statistics} lie mainly between $1.5d$ and slightly above $2d$, consistent with this classical breakup scale despite the use of MHz ultrasound excitation. The data should be interpreted as distributions of optically resolved millimeter-scale objects sampled by the virtual gate, not as complete spray distributions. This distinction is important because classical ultrasonic atomization and modern ultrasonic-nebulizer studies show that MHz capillary-wave atomization can generate much smaller droplets than what can be resolved at the present magnification \citep{Lang1962,Kooij2019}. Thus, any sub-resolution mist produced near the vibrating nozzle or droplet field is outside the quantitative size range of Fig.~\ref{fig:droplet_statistics}. 

In the fixed-flow voltage sweep at $\mathrm{We}=5.56$ and $\mathrm{Oh}=0.0038$ (Fig.~\ref{fig:droplet_statistics}a), the resolved droplet distribution changes only modestly at low and intermediate $u^*$, but high forcing produces distribution-level shifts relative to the unforced case. The largest shifts occur at $u^*=0.54$, 0.64, and 0.94, which are also in the actuation range where the breakup-length data show strong acoustic modification. The median visible droplet diameter does not decrease monotonically with acoustic forcing. Instead, the high-$u^*$ cases contain a visibly larger upper portion of the resolved distribution.

The flow-rate sweep provides the complementary comparison at fixed acoustic drive (Fig.~\ref{fig:droplet_statistics}b). At each Weber number, the $25~\mathrm{V_{pp}}$ case is compared with the corresponding unforced case using the same region of interest (ROI) and gate for that flow rate. Acoustic actuation shifts the resolved diameter distribution upward for most of the Rayleigh-regime conditions, with differences from $\mathrm{We}\approx 6$ through $\mathrm{We}\approx 39$. At $\mathrm{We}\approx 50$, where the breakup-length reduction also weakens as the jet flow speed increases, reducing the relative amplitude of the acoustic forcing over the intrinsic velocity of the jet, causing the on/off difference in the resolved diameter distribution to be less pronounced. The droplet-size data therefore reinforce the same physical picture obtained from $L_b^*$: MHz excitation strongly modifies the breakup process at low and moderate $\mathrm{We}$, but its relative influence weakens as jet inertia increases.

\section{Conclusions}
We have shown that MHz-order acoustic excitation applied directly at the nozzle exit can substantially modify Rayleigh-regime liquid jet breakup in quiescent air. A through-hole thickness-mode lithium niobate transducer operating near \SI{6.7}{MHz} generated visible interfacial corrugations and shifted the breakup point upstream. By analyzing high-speed shadowgraphy data with an autocorrelation-based event-selection method, we separated instantaneous jet-length fluctuations from statistically independent breakup events and quantified the normalized breakup length $L_b/d$.

The acoustic response depends on forcing amplitude, jet velocity, viscosity, and base wetting. For water at low Weber number, increasing acoustic forcing produced a threshold-like reduction in $L_b/d$, followed by saturation and partial recovery at large We. A reduced-order threshold correlation was fit to the water amplitude and velocity sweeps and captured the sudden decrease in $L_b/d$ and the loss of acoustic efficacy near $\mathrm{We}\simeq 50$. Because the correlation was not validated against independent conditions, it should be interpreted as a compact representation of the water data rather than as a predictive theory. At fixed acoustic voltage, the reduction weakened as $\mathrm{We}$ increased because the normalized forcing $u^*=\dot{\xi}_p/u_0$ decreased, the disturbance had less residence time to grow, and the wetted-base coupling path weakened near the wind-influenced transition.

The glycerol--water mixtures did not follow the water-threshold behavior. Instead, the higher-viscosity, higher-Ohnesorge-number fluids shortened immediately at low forcing, indicating a different acoustic-coupling and breakup regime. Increasing viscosity lengthened the unforced jet but did not suppress acoustic control over the tested range. Base-wetting experiments showed that a wetted nozzle base independently reduces $L_b/d$, confirming that wetting-mediated coupling is an important part of the mechanism.

Resolved droplet tracking showed that the acoustic field also changes the visible detached droplet population. For water, high normalized forcing and fixed-voltage excitation across most of the Rayleigh-regime flow sweep shifted the distributions of gate-crossing $D_{\mathrm{eq}}^*$ relative to the corresponding controls. These data do not show a simple monotonic reduction in resolved droplet diameter; rather, they indicate that MHz excitation changes the breakup pathway and the distribution of visible millimeter-scale fragments. Droplets below the three-pixel optical cutoff and any micron to submicron-scale atomized mist remain unresolved in the present measurements.

 Future work should quantify the full spray-size distribution using higher-magnification imaging, laser diffraction, or phase-Doppler diagnostics, measure internal flow or streaming near the nozzle base, and independently vary $\mathrm{We}$, $\mathrm{Oh}$, and $u^*$ to separate inertia, viscous damping, and acoustic coupling effects. These results establish MHz-order acoustics as a compact and tunable approach for controlling free-surface jet breakup without requiring electrical, magnetic, or chemical modification of the fluid.

\section*{Authors' contributions}
K.N., K.A., and J.F. conceived the study. K.N. designed and fabricated the nozzle--transducer assembly, performed the experiments, analyzed the high-speed-imaging data, and prepared the initial manuscript draft. J.F. supervised the project, developed the reduced-order data correlation, and contributed to interpretation of the acoustic and fluid-mechanics results. K.N., K.A., and J.F. revised the manuscript. All authors reviewed and approved the final manuscript.
% chedk the format for eip Amihai Horesh: Investigation (equal); Methodology (equal); Writing –
% original draft (equal). William Connacher: Formal analysis (equal);
% Investigation (equal); Methodology (equal); Resources (equal);
% Visualization (lead); Writing – original draft (lead). James Friend:
% Conceptualization (lead); Formal analysis (lead); Funding acquisition
% (lead); Investigation (lead); Project administration (lead); Resources
% (lead); Supervision (lead); Writing – review & editing (lead).

\section*{Acknowledgments}
Fabrication was performed in part at the San Diego Nanotechnology Infrastructure (SDNI) of UCSD, a member of the National Nanotechnology Coordinated Infrastructure, which is supported by the National Science Foundation (Grant ECCS–1542148). The authors wish to express their gratitude to Professor Nigel Goldenfeld (UCSD) and Professor Philip Marston (Univ.\ Washington) for discussions regarding this work.

\section*{Funding}

The authors declare that financial support was received for the research, authorship, and/or publication of this article. This work was supported by funding from the Office of Naval Research, United States, under grant No. 13423461 to J.F. and K.A. Additional support for K.N. was provided by the STARS Fellowship.

\section*{Data availability}
All data associated with this manuscript is available online at \href{https://doi.org/10.6075/J02N52F5}{https://doi.org/10.6075/J02N52F5}.

\section*{Competing interests}
The authors declare that they have no competing interests.

%% Create the reference section using BibTeX.
\bibliography{thruhole}

\end{document}